\documentclass[a4paper]{jpconf}
\pdfoutput=1
\usepackage{graphicx}
\usepackage{tabularx}
\usepackage{color}
\usepackage{multicol}
\usepackage{siunitx} 
\usepackage{subfigure}
\usepackage{amsmath, amssymb}

\begin{document}

\twocolumn
[
\title{Probing domain walls in cylindrical magnetic nanowires with electron holography} 

\author{M Sta\v{n}o\,$^{1}$, S Jamet\,$^{1}$, J C Toussaint\,$^{1}$, S Bochmann\,$^{2}$, J Bachmann\,$^{2}$, A Masseboeuf\,$^{3}$, C Gatel\,$^{3}$ and O Fruchart\,$^{1,4}$}

\address{$^{1}$ Univ. Grenoble Alpes - CNRS, Inst NEEL, F-38000 Grenoble, France}
\address{$^{2}$ Univ. Erlangen, D-91054 Erlangen, Germany}
\address{$^{3}$ CNRS, CEMES, F-31400 Toulouse, France}
\address{$^{4}$ Univ. Grenoble Alpes - CNRS - CEA, SPINTEC, F-38000 Grenoble, France}

\ead{michal.stano@neel.cnrs.fr}

\begin{abstract}
We probe magnetic domain walls in cylindrical soft magnetic nanowires using electron holography. 
We detail the modelling of expected contrast for both transverse and Bloch point domain walls and provide comparison with experimental observations performed on NiCo nanowires, involving also both magnetic and electrostatic contribution to the electron holography map. This allows the fast determination of the domain wall type without the need for uneasy and time-consuming experimental removal of the electrostatic contribution. Finally, we describe and implement a new efficient algorithm for calculating the magnetic contrast. 
\end{abstract}
] 
\section{Introduction}

  In this study we were interested in imaging magnetic domain walls (DWs) in rather soft magnetic cylindrical nanowires by means of electron holography~\cite{lichte2008e-holo}. 
Aside from fundamental research interest, IBM proposed a concept of three-dimensional solid-state magnetic race-track memory based on shifting domain walls in magnetic nanowires~\cite{Parkin2008}. In our work, the focus was on creating and localizing a DW and identifying its type with the help of numerical simulations. 

Biziere et al.~\cite{Biziere2013} already applied electron holography to nickel nanowires. They identified both a Transverse Wall (TW) and a Transverse Wall with significant curling for nanowires with diameters 55\,nm and 85\,nm, respectively. For larger diameters another DW may also be found, with lower energy than TWs. It is the so called Bloch Point Wall (BPW), characterized by a purely orthoradial curling. We have recently provided the first experimental observation of the BPW~\cite{DaCol2014} by polarized synchrotron X-rays. 

Here we considered Ni$_{60}$Co$_{40}$ nanowires with diameter 100-150\,nm, so as to favor BPWs. Nevertheless, we observed only TWs, following nucleation with a transverse field. We show how it is necessary to consider the electrostatic potential, as well as the azimuthal degree of freedom of the transverse wall, to formally reproduce and thus identify the experimental contrast. 

\section{Electron holography}
Electron holography enables us to determine quantitatively both amplitude and phase of the electron wave that passes through a thin sample, thanks to interference and phase analysis on fringes frequency. 
The phase shift of the electron wave with respect to vacuum is affected by both electrostatic and magnetic potentials. 
 Note that only in-plane (perpendicular to the electron beam) magnetic induction components contribute to the magnetic phase shift. Therefore the sample needs to be tilted to get information about the remaining induction component.

\section{Experimental procedures}


Electron holography imaging was conducted on the I$^2$TEM microscope (Lorentz mode; using a transfer lens corrector as imaging lenses). A double biprism was used to enlarge field of view and remove Fresnel fringes. We considered Ni$_{60}$Co$_{40}$ nanowires with diameter 100-150\,nm, length at least 20\,\si{\micro}m, terminated at either end with 200nm-diameter parts to deter the DW annihilation. The nanowires were dispersed on a Cu grid with a lacey carbon thin film. 

The objective lenses of the microscope were used to apply field of 1\,T transverse to the wire axis to nucleate a DW and then switched off to allow imaging at remanence. 
In some cases the electrostatic contribution to the phase was experimentally removed via subtracting images with a flipped sample. 
    
\section{Simulation} 

We calculated expected phase maps based on two micromagnetic configurations with a domain wall - either TW or BPW. This can be done for different wire tilts and domain wall orientation (azimuth for the TW) for both magnetic and electrostatic contributions to the electron phase shift.

First we compute the relaxed 3D micromagnetic configuration of the domain wall in the wire using FeeLLGood~\cite{feeLLGood-web-only}, a home-built code based on the temporal integration of the Landau-Lifshitz-Gilbert equation in a finite-element scheme, i.e. using tetrahedra cells. Only exchange and dipolar interactions were taken into account (thus neglecting magnetocrystalline anisotropy) with the following parameters: saturation polarization $\mu_0 \mathrm{M_{\rm s}}$ = 1\,T and exchange stiffness $A = 10^{-11}$\,J/m. The tetrahedron size was 5\,nm or smaller. We consider a wire with the length sufficiently longer than the DW width, however much shorter than in the experiment. Magnetic charges are removed at either end, thus mimicking an infinitely long wire. 

The relaxed micromagnetic configuration is then used as an input for the calculation of the electron holography phase maps. Our technique combines both computational efficiency and precision which is uncommon for a numerical method, mainly due to the transformation of a 3D problem to a 2D one. Instead of performing the calculation in the whole ($x$,$y$,$z$) space like in~\cite{Biziere2013}, the magnetic phase shift is proportional to the $z$-component of the magnetic vector potential produced by a 2D equivalent magnetic system whose in-plane magnetization components are given by the path integral $\int  dz \; (M_{y}, -M_{x})$, where $z$ denotes the beam direction. The latter expression corresponds to the integrated in-plane magnetization of the 3D original system rotated by $90$\,$^{\circ}$ around the beam axis.

We assume the electrostatic part to be proportional to the sample thickness and to a volume-averaged Mean Inner Potential of 22\,V with the constant of proportionality $\mathrm{c}_{300\mathrm{\,keV}}=6.5262\cdot 10^{6}\,\mathrm{\,rad/(V\cdot m)}$~\cite{lichte2008e-holo}.

\begin{figure*}
  \begin{minipage}{0.89\linewidth}
	\centering 
		\includegraphics[width=4.2cm]{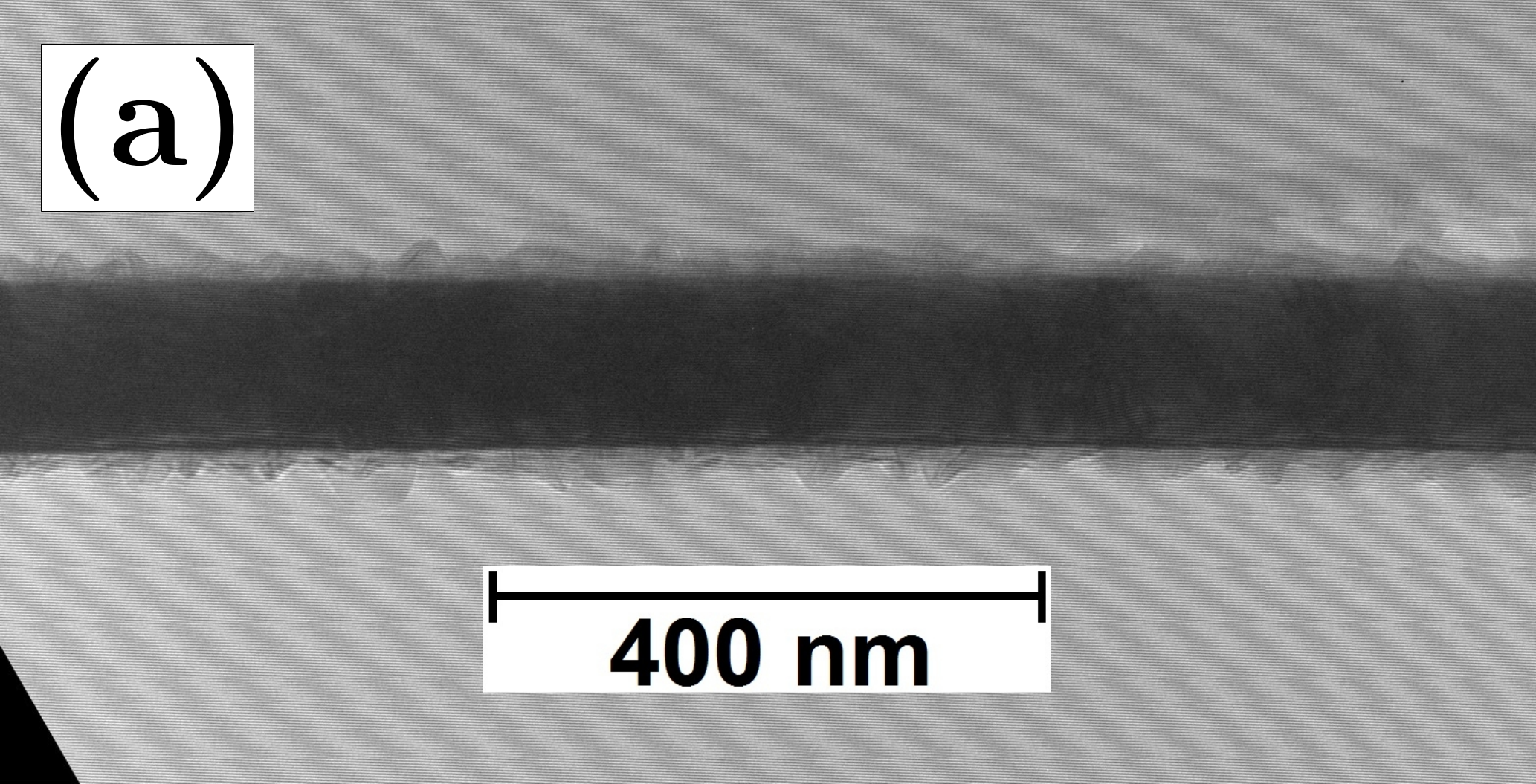} 
		\includegraphics[width=4.2cm]{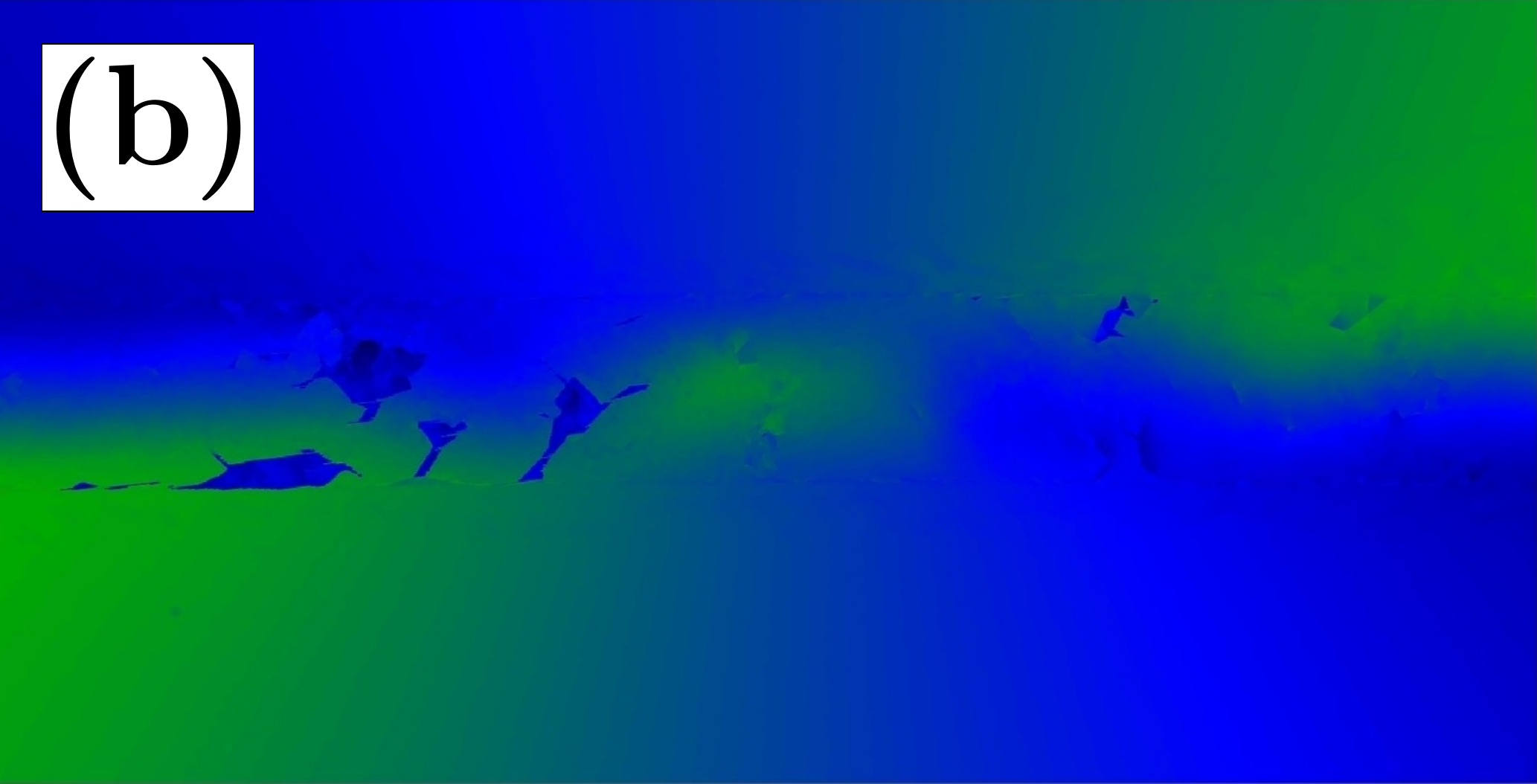} 
		\includegraphics[width=4.2cm]{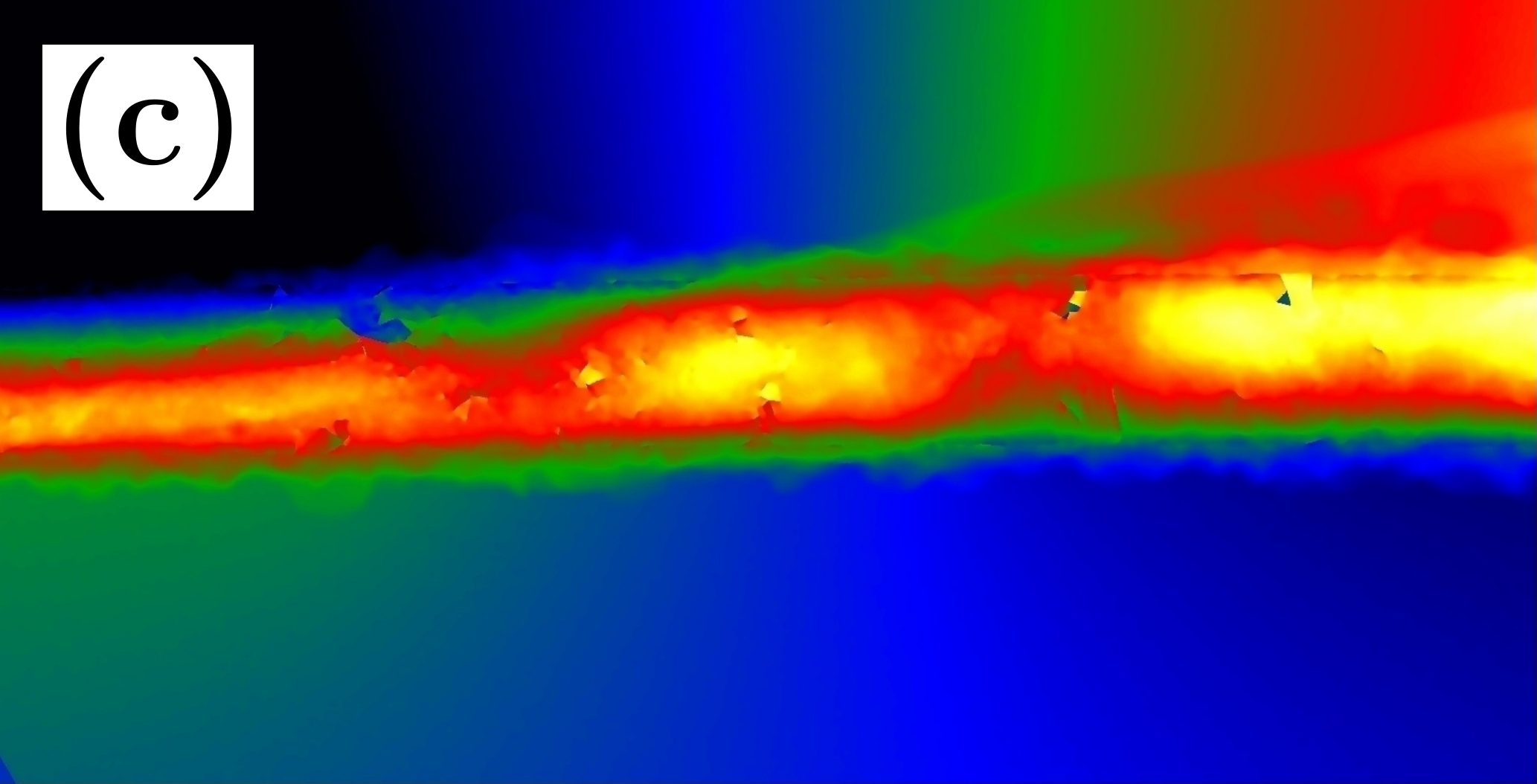} \\ \vspace{1mm} 
		\includegraphics[width=4.2cm]{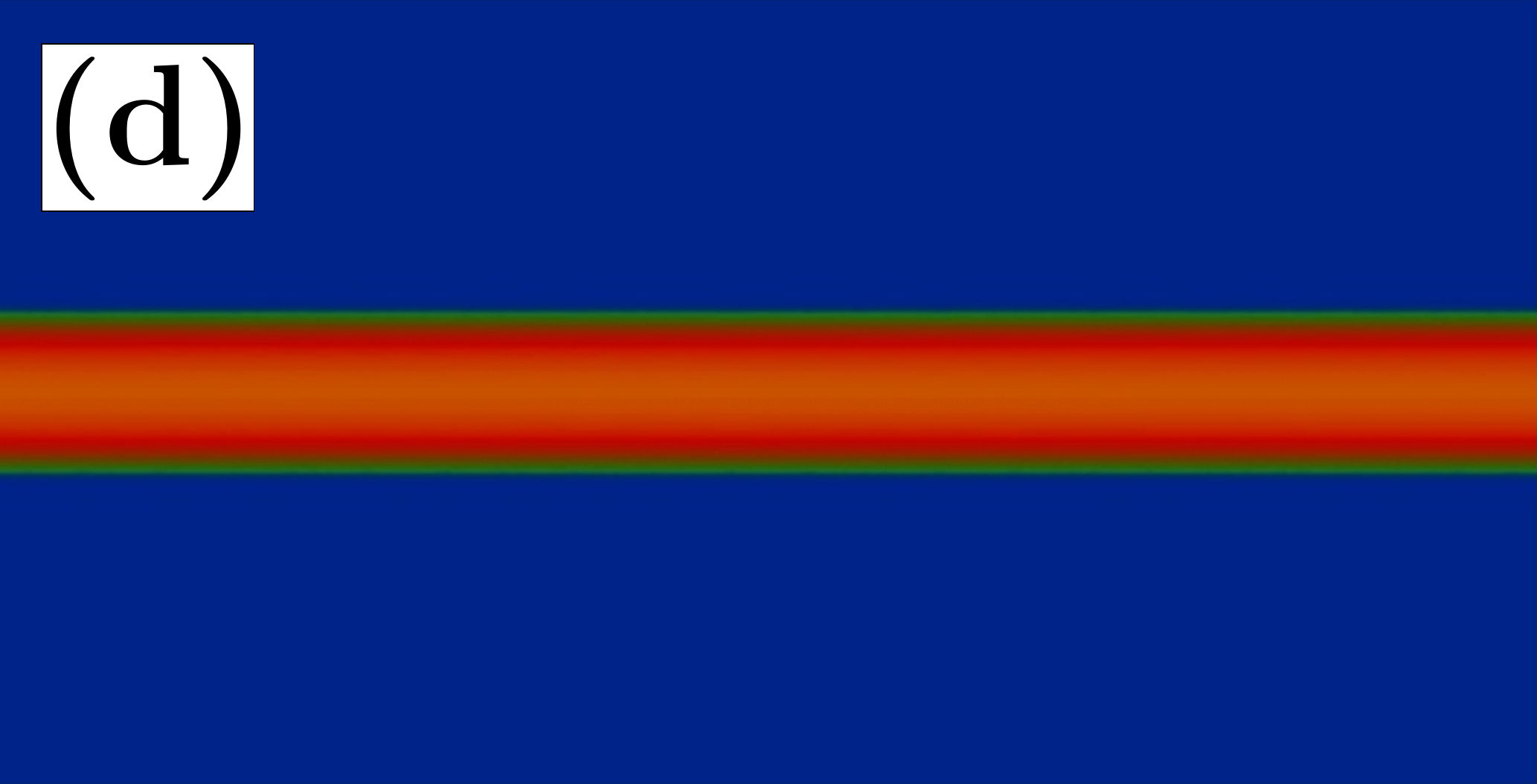} 
		\includegraphics[width=4.2cm]{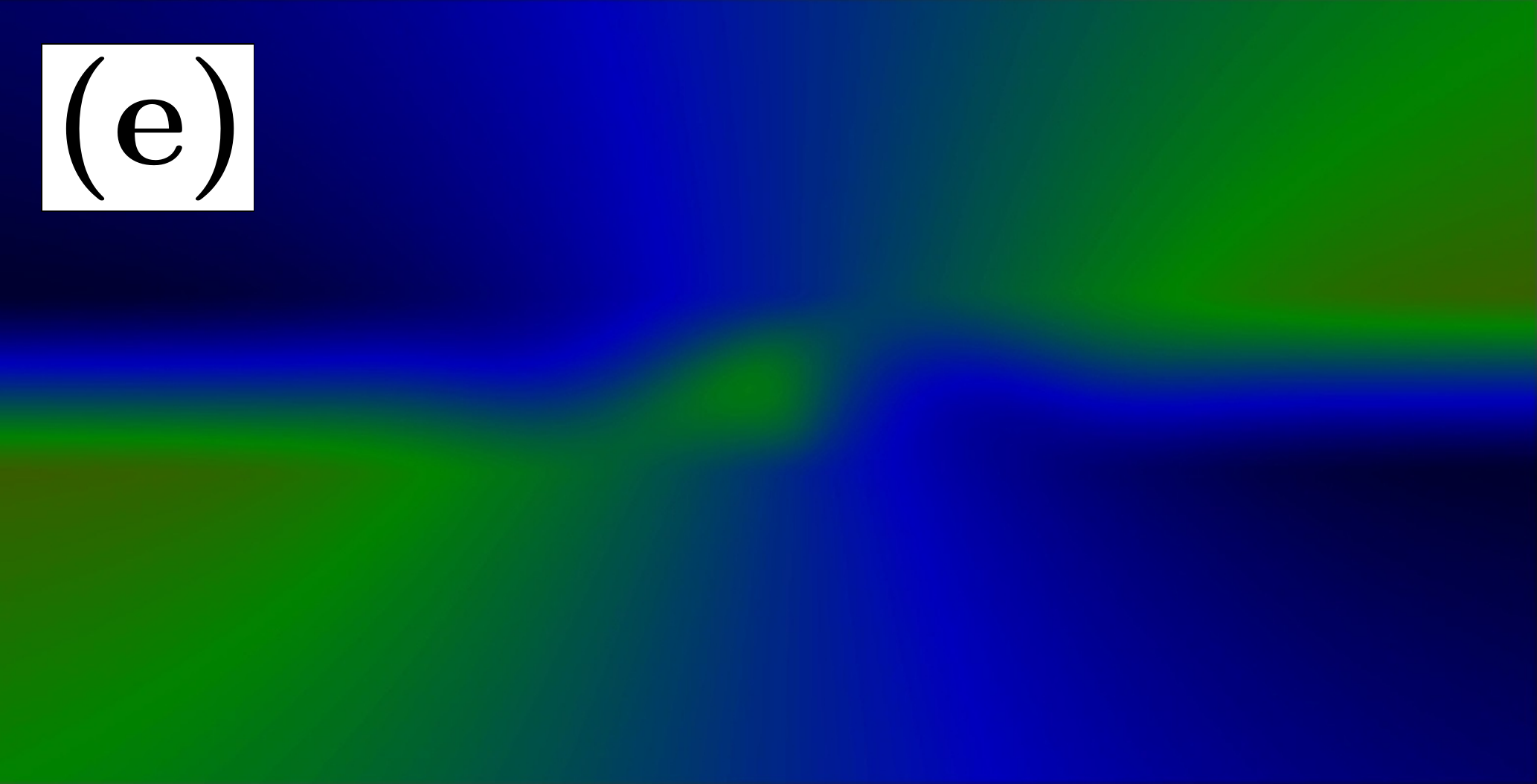} 
		\includegraphics[width=4.2cm]{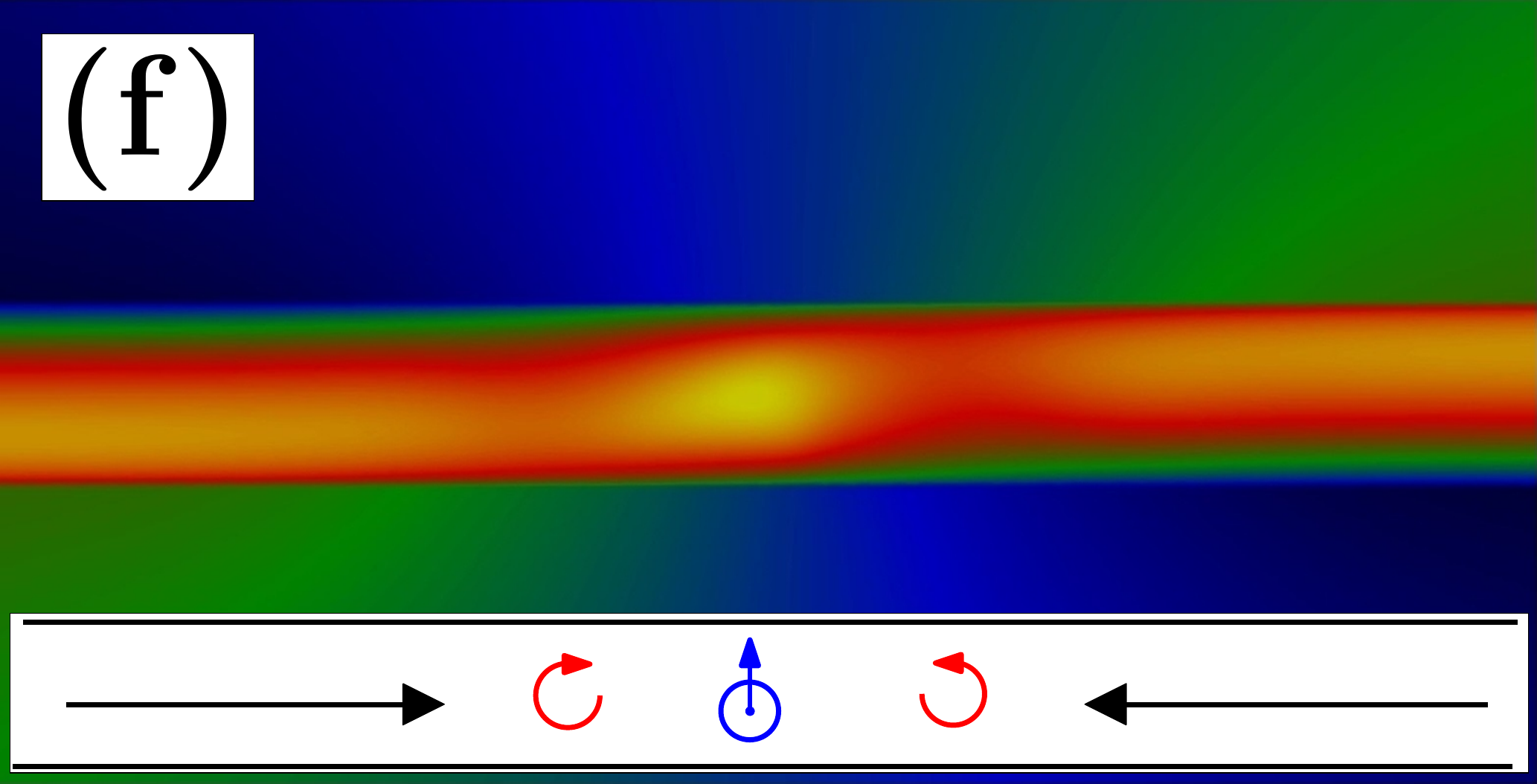} \\  
	\end{minipage} \hspace{-5mm}
  \begin{minipage}{0.1\linewidth}
	\centering
	\includegraphics[height=4.4cm]{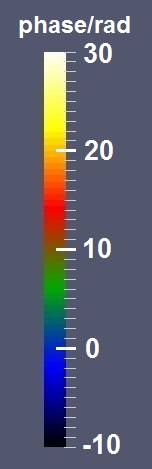}
	\end{minipage}
		\vspace{-1mm}
		\caption{Electron holography. (a) hologram (wire+interference fringes); experimental phase maps: (b) magnetic only and (c) electrostatic+magnetic. Simulated phase: (d) electrostatic only, (e) magnetic only and (f) both contributions with a scheme of the magnetization pattern.}
	\label{img_holo+MIP}
		\vspace{-2mm}
\end{figure*}

\section{Results and discussion}

The electron phase shift maps for a wire with a domain wall (reconstructed from the electron hologram) and their numerical modelling are summarized in Fig.~\ref{img_holo+MIP}. The electrostatic part which may make the DW identification more difficult can be either removed experimentally or included in the simulations. The first option paves the way towards quantitative matching of the DW pattern but it is time consuming and not straightforward. On the other hand, even with a rather simple model for the electrostatic part we can reproduce the experiment numerically without this nuisance.
The experimental magnetic phase shift is slightly smaller than in the simulations; we attribute this to a reduced saturation magnetization due to surface oxidation. Quantitative agreement for both magnetic and electrostatic contributions is possible yet difficult due to the model simplicity and electrostatic contributions arising from an inhomogeneous supporting carbon film, defects and crust of impurities on the wire [Fig.~\ref{img_holo+MIP} (a)]. 

\begin{figure}[ht]
\begin{minipage}{0.7\linewidth}
	\centering
		 
		\includegraphics[height=2.4cm]{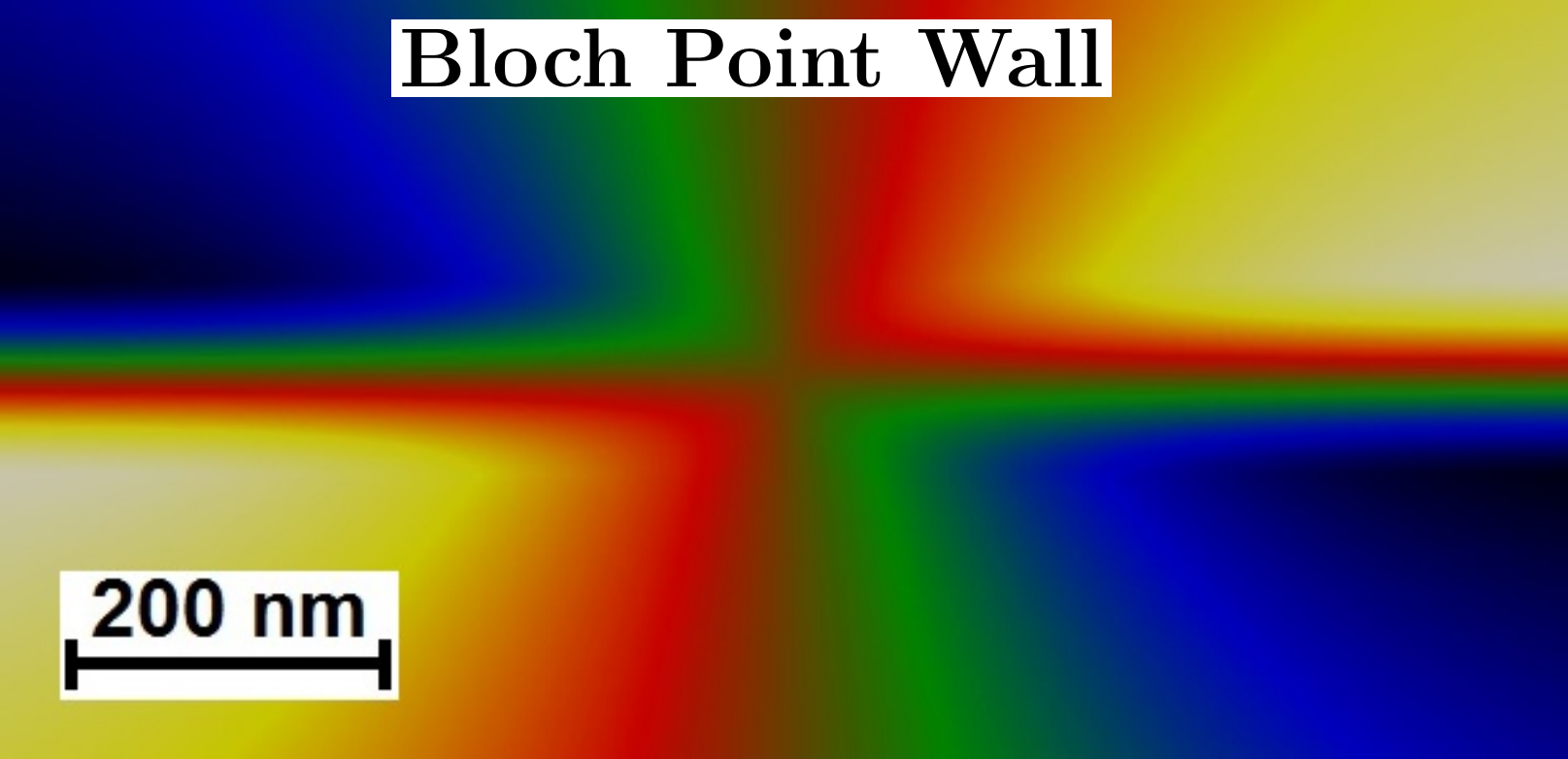} \\ \vspace{0.5mm}
		\includegraphics[height=2.4cm]{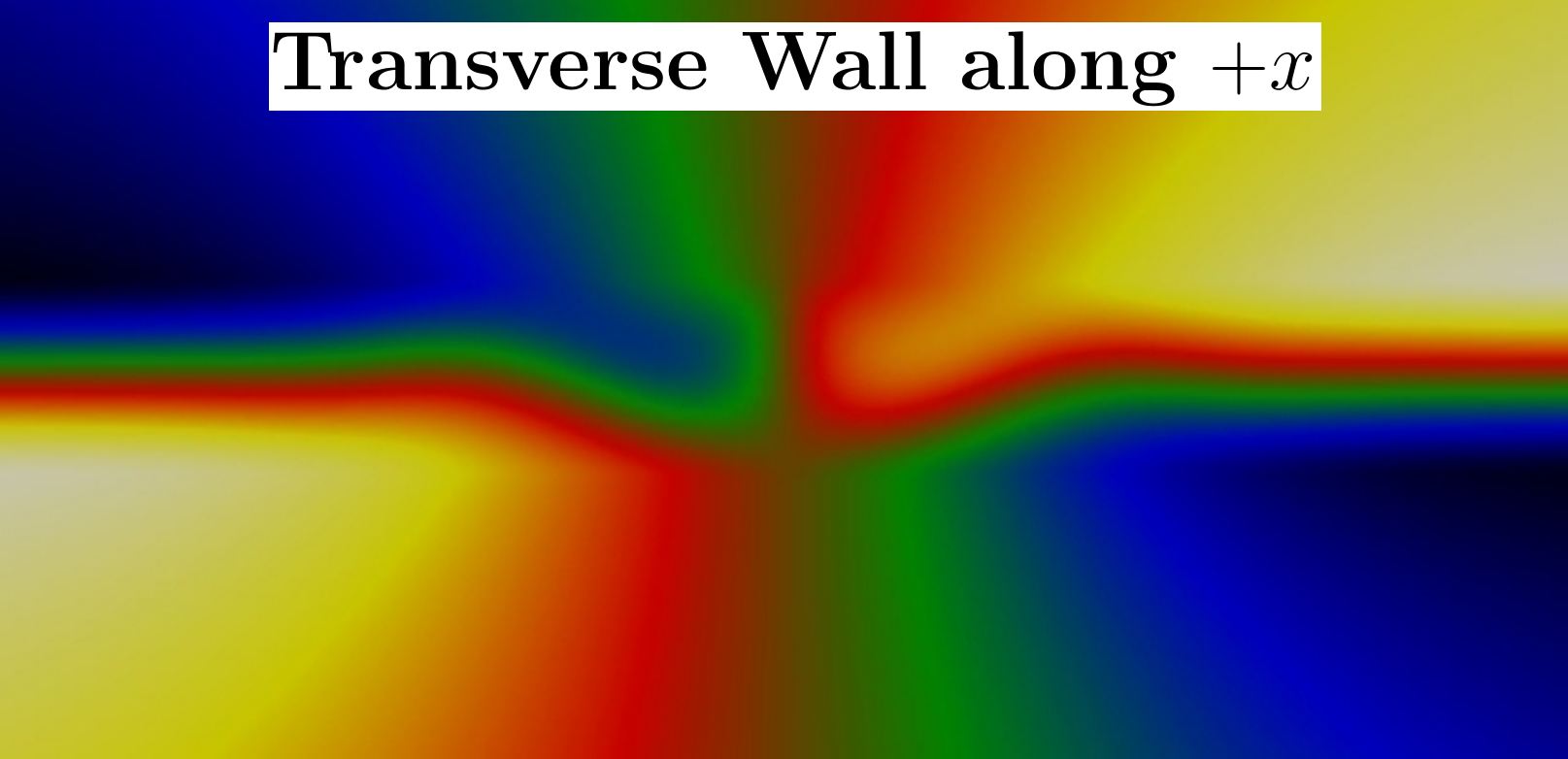} \\ \vspace{0.5mm}
		\includegraphics[height=2.4cm]{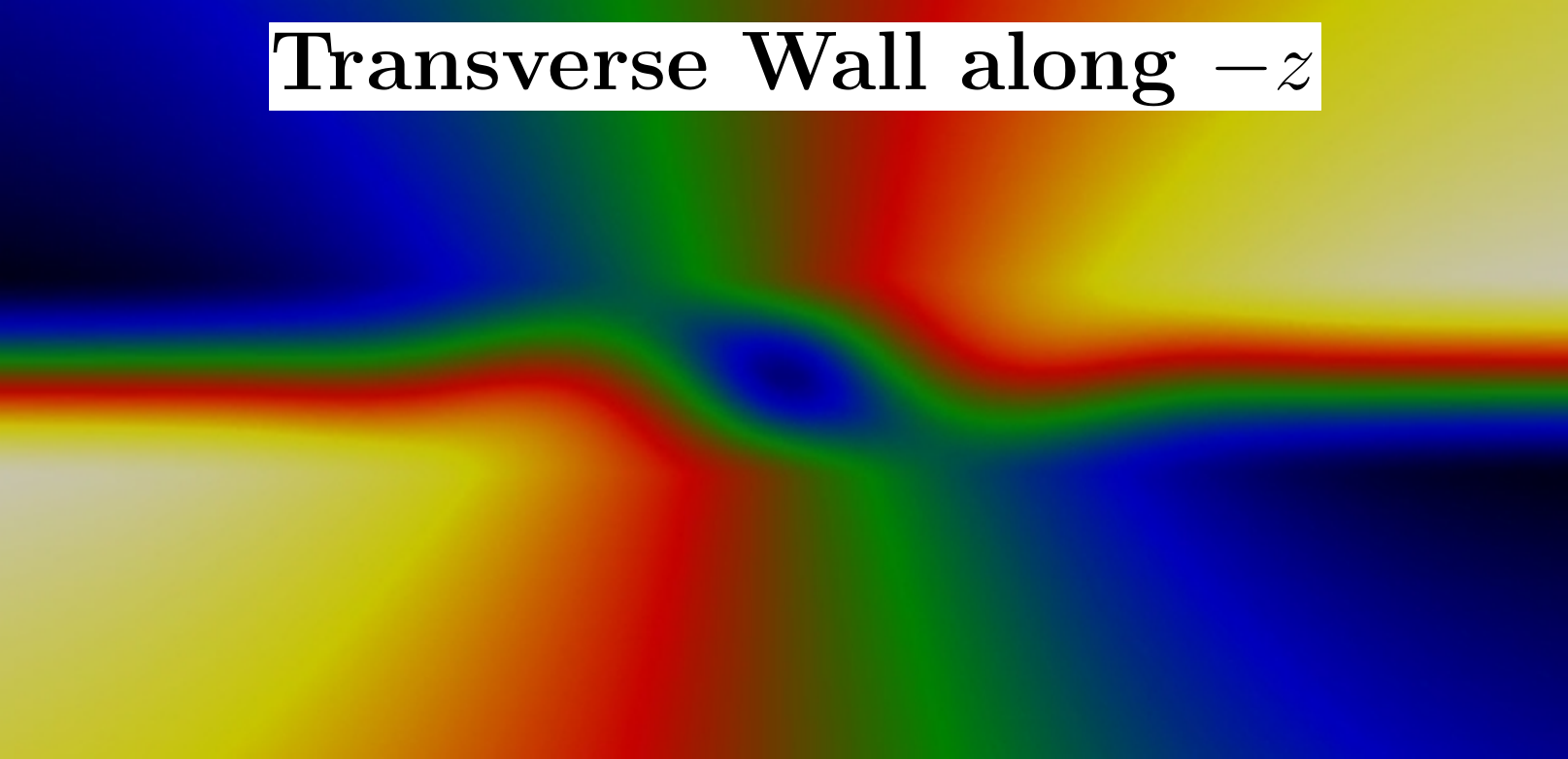} \\ \vspace{0.5mm} 
		\includegraphics[height=2.4cm]{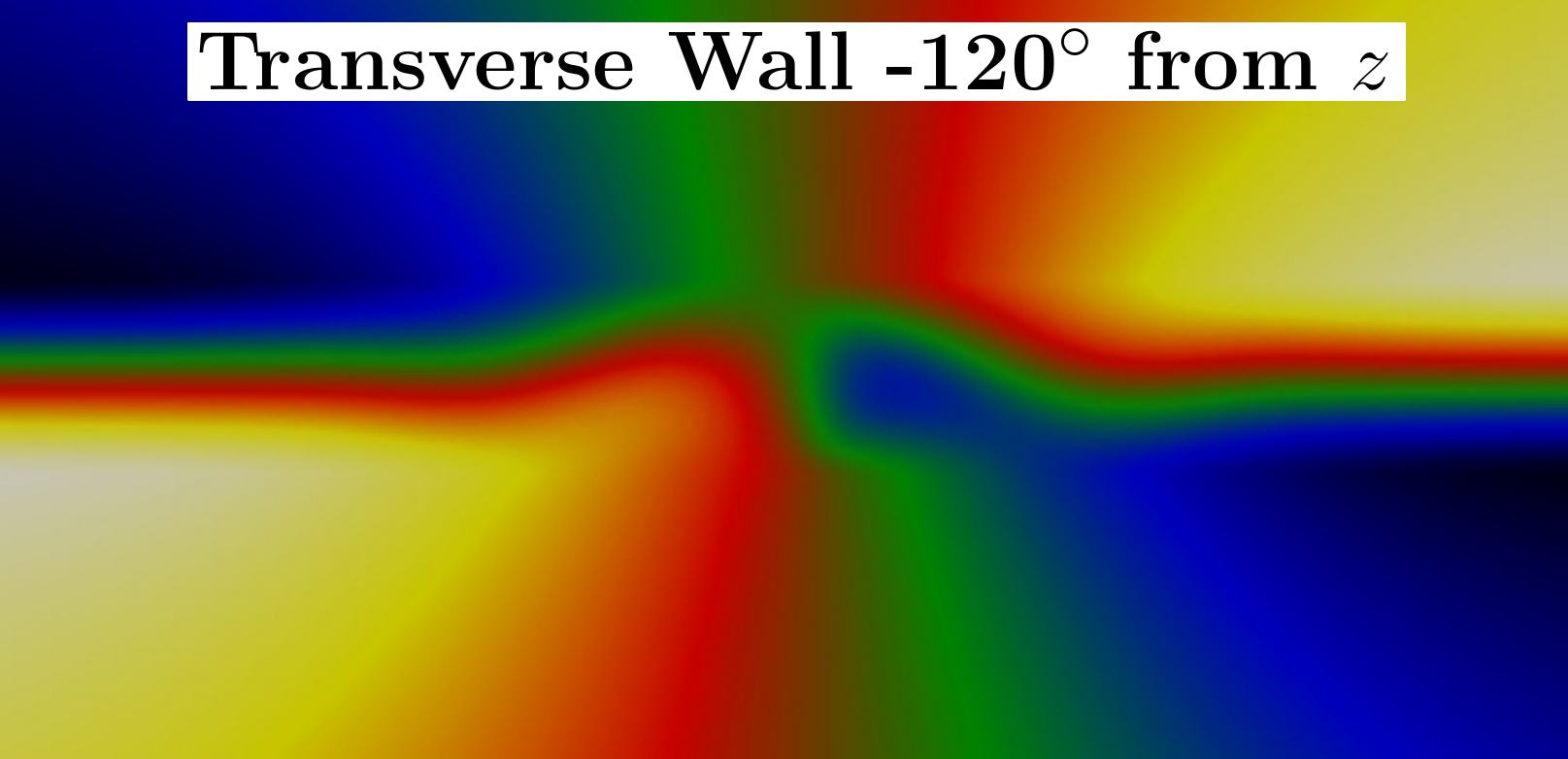} \\ 
		\includegraphics[height=2.4cm]{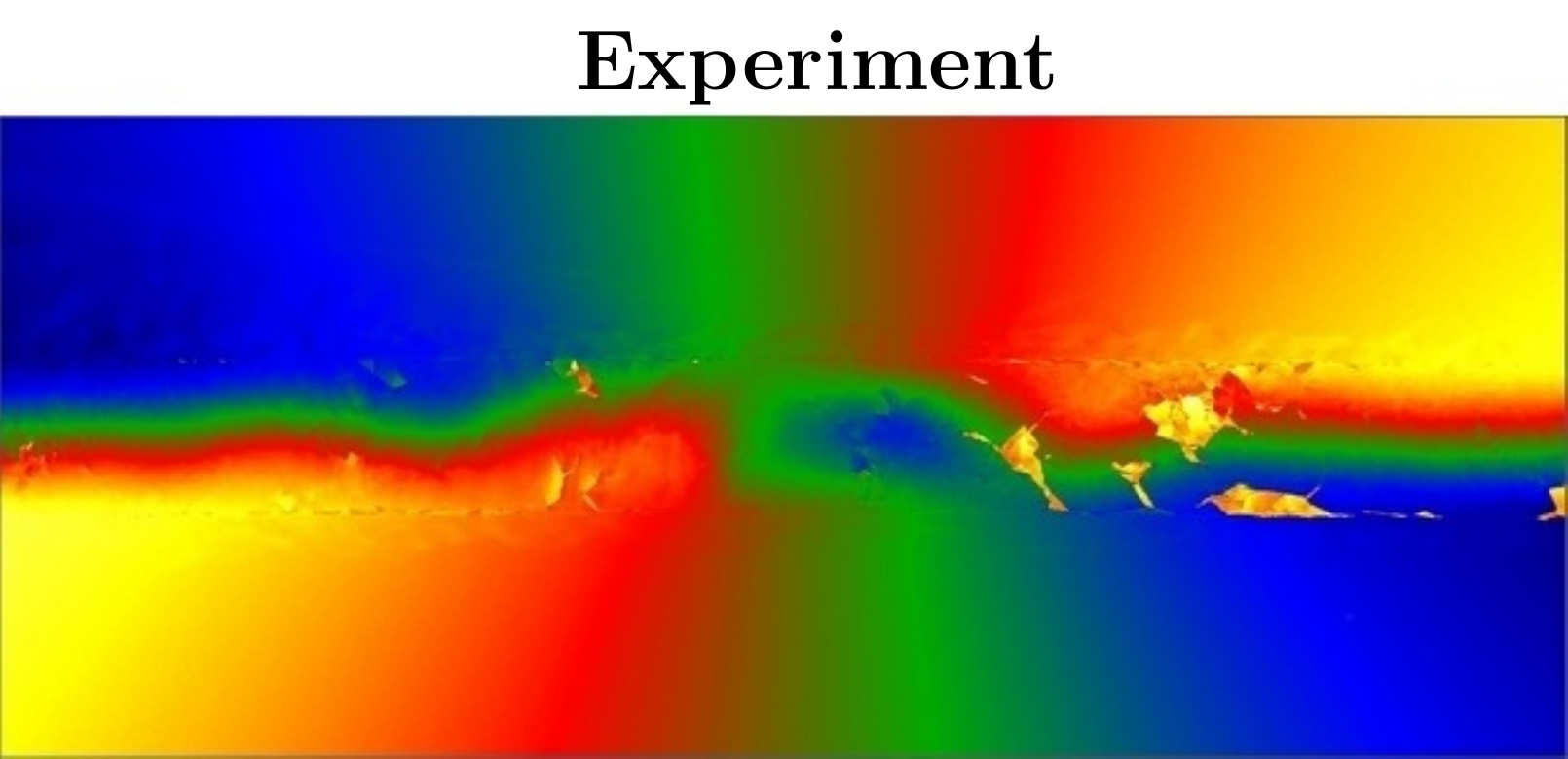} 
\end{minipage}
\begin{minipage}{0.25\linewidth}
\centering
		\includegraphics[width=2cm]{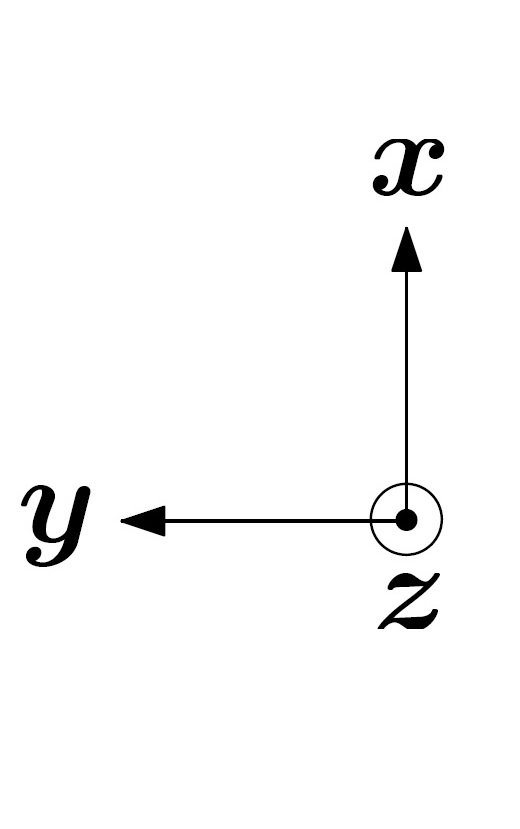}
		beam $\boldsymbol{\parallel z}$
		\\
		\bigskip
		\medskip
		\includegraphics[width=2cm]{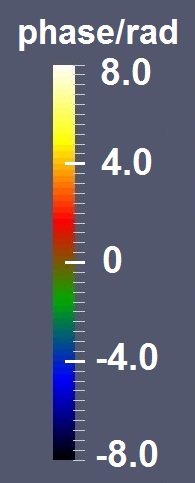} 
\end{minipage}
		\caption{Magnetic phase-shift maps for both DWs and different azimuth of the TW. Here transverse wall with azimuth of -120\,$^{\circ}$ (vs $z$) was identified.}
		\label{img_holo_mag_only}
		\vspace{-2mm}
\end{figure}

In all our experiments we identified the transverse wall with various orientation (azimuth). Fig.~\ref{img_holo_mag_only} shows that depending on this azimuth, the same DW produces different magnetic phase maps. We can also distinguish BWP and TW along the beam direction thanks to an additional curling that has opposite sense at either side of the TW - this has been already reported in~\cite{Biziere2013}. For the BPW the curling sense is the same on both sides. Note that different magnetic configurations can produce very similar phase maps - e.g. BPW and ideal TW without curling pointing along the beam (exist only for small diameters~\cite{DaCol2014} - not our case). Therefore one should acquire images for at least few tilts to avoid the misinterpretation. 

\section{Conclusion} 

Electron holography is a great tool for fast determination of a domain wall presence and together with simulations enables identification of the DW and determination of its structure. 
Our new efficient algorithm for simulations of the electron holography phase maps shows very good agreement with the experiment. Both magnetic and electrostatic contributions to the phase shift can be modelled for various magnetic nano/micro structures as demonstrated here on cylindrical nanowires.

\section*{Acknowledgements}
This project has received funding from the European Union Seventh Framework Programme (FP7/2007-2013) under grant agreement n$^{\circ}$~309589 (M3d). The experiments were conducted via the French research federation METSA. 

\section*{References}
\providecommand{\newblock}{}


\end{document}